\documentstyle[11pt]{article}

\newcommand{\be}{\begin{equation}}
\newcommand{\ee}{\end{equation}}
\newcommand{\ba}{\begin{array}}
\newcommand{\ea}{\end{array}}
\newcommand{\bc}{\begin{center}}
\newcommand{\ec}{\end{center}}
\newcommand{\bi}{\begin{itemize}}
\newcommand{\ei}{\end{itemize}}

\newcommand{\disregard}[1]{{}}

\def\bild#1\over#2{\mathrel{\mathop{\kern0pt #1}\limits_{#2}}}

\begin{document}

{\centerline {\bf On $\delta$ perturbative interactions in the
Aharonov-Bohm and Anyon Models\rm}}
\vskip 1cm
{\centerline {\bf St\a'ephane OUVRY \rm
\footnote{\it
 electronic e-mail: OUVRY@FRCPN11}}}

{\centerline {Division de Physique Th\'eorique \footnote{\it Unit\a'e de
Recherche  des
Universit\a'es Paris 11 et Paris 6 associ\a'ee au CNRS},  IPN,
  Orsay Fr-91406 }}

{\centerline{and}}

{\centerline{ L.P.T.P.E., Universit\'e Pierre et Marie Curie, 4 Place Jussieu,
75252, Paris Cedex 05}}

\vskip 1cm

Abstract - One discusses the validity and equivalence of various perturbative
approaches for the Aharonov -Bohm  and Anyon models.
\vskip 1cm

\vfill\eject

Let us consider the paradigm problem [1] of a charged planar particle $ e$ in
the
gauge potential of a singular flux tube $\phi$ located at the origin
\be\label{1} H^{\alpha}={1\over2m_o}\left(\vec p-\alpha{\vec k\times\vec r\over
r^2}
            \right)^2 \ee
$\vec k $ is the unit vector perpendicular to the plane;
$\alpha=e\phi/2\pi$ the Aharonov-Bohm (A-B) or statistical parameter (in the
latter case
(\ref{1}) should be considered as the relative 2-anyon\footnote{ For the
time being one concentrates on the 2-anyon case, where the essence of the
perturbative problem and its solution can be easily described.
Generalization to the $N$-anyon case will be displayed afterwards.}
Hamiltonian [2] with
$m_o\to m_o/2$). In what follows,
$\alpha$ will always be considered close to zero.
The spectrum is continuous $E_{km}^{\alpha}=k^2/2m_o$
with normalized states in the continuum
\be\label{2} \psi_{km}^{\alpha}(r,\theta)=
            \sqrt{k\over2\pi}e^{im\theta}J_{|m-\alpha|}(kr) \ee
We remark
that
${\sqrt{k/2\pi}}\exp(im\theta)J_{-|m-\alpha|}(kr)$ is also a solution,
but not normalizable, except for the s- wave $m=0$. In this case,
one
should in principle consider a
linear combination
of both solutions $J_{|\alpha|}$ and $J_{-|\alpha|}$. This is paramount to
self-adjoint
extension considerations [3] which are, however,
 forbidden if we assume that the eigenstates should vanish at the
origin ($J_{-|\alpha|}\to r^{-|\alpha|}$
when $r\to 0$), meaning unpenetrable solenoids in the A-B model [4]
or
exclusion of  the diagonal of the  configuration space in the anyon model.

The non analytical behavior of the $m=0$ eigenstates in
$\alpha$
is a clear indication of the failure of a standard
perturbative analysis. This is due to the fact that the s-wave ($m=0$)
unperturbed
Hilbert space (bosonic Hilbert space in the anyon context), is not adapted to
the domain of
definition of the Hamiltonian, since the unperturbed $m=0$ eigenstates do
not vanish at
the
origin to the contrary of the exact $m=0$ eigenstates which do vanish as
$r^{|\alpha|}$ when $r\to 0$.

Clearly, perturbation has to be singular, and indeed logarithmic
divergences show up
in the
computation of the matrix elements of the Hamiltonian (\ref{1}),
due to the singular nature
of the $1/{r^2}$ interaction at the origin (coinciding points).

For pedagogical reasons, let us display explicitely
how these perturbative singularities
materialize, by confining\footnote{
Here, one treats short range perturbative singularities, where
the harmonic regularization is
purely a matter of convenience.  On the other hand, harmonic
regularization is crucial
when one is interested in the
thermodynamic limit [5], which is not the issue adressed in the present
analysis.}
the system in a harmonic potential ${1\over 2} m_o\omega^2r^2$.
One has the exact eigenstates
\be \label{10}\psi_{nm}^{\alpha}= {\cal N}_{n,|m-\alpha|}e^{im\theta}e^{-\xi/2}
\xi^{\vert  m-\alpha
\vert/2}L_n^{\vert  m-\alpha
\vert}(\xi)\ee
 ($\xi=m\omega r^2/2$;
 ${\cal N}_{n|m-\alpha|}$ is a normalization factor) and the discrete spectrum
\be \label{11} E_{nm}^{\alpha}= (2n+\vert  m-\alpha \vert +1)\omega\ee
Perturbing the free Hamiltonian by a small A-B (anyonic) interaction, one
estimates
the matrix element
\be \label{12}\alpha <\psi_{nm}^0
(r, \theta)\vert {2i\over r^2}\partial_{\theta} \vert
\psi_{nm}^{0} (r, \theta )>\ee
where $\psi_{nm}^0
(r, \theta )$ stands for the standard zeroth order wave function.
The result is
 \be \label{13}\ba {ll} -{m\over \vert m\vert}\alpha\omega \  \ &{\rm for}\
m\ne 0 , \\
  0 \ \ \ &{\rm for}\  m=0.\ea \ee
 There is obviously no contribution
 to the s-wave states.
Perturbation theory makes sense
only when all perturbative corrections to the zeroth order spectrum are finite.
However, if
\be \label{14}
\alpha^2<\psi_{nm}^0
(r, \theta)\vert {1\over r^2} \vert
\psi_{nm}^0 (r, \theta)>\simeq
\alpha ^2\int d^2\vec r  {r^{2\vert m\vert}\over r^2}
\to \alpha^2 \int dr  r^{2\vert m\vert -1},\ee
is properly defined
if $m \ne 0$,  it
is logarithmically divergent if $m =0$.
Thus one concludes that perturbation theory breaks down for the
$m=0$ states, i.e. for Bose statistics in the anyon context.

They are several approaches to solve this problem :

i) the non hermitian Hamiltonian perturbative approach [6]

ii) the perturbative approach around a good Hilbert space [7]

iii) the perturbative approach
around the standard Hilbert space but with the addition of a repulsive
$\delta$ interaction [8].

After a short review of i), ii) and iii), I will show how and why they
are equivalent.

i) the non hermitian Hamiltonian perturbative approach [6] :

As already
noticed when discussing possible self-adjoint extansions,
the $m=0 $  states have to be treated with a  particular care.
It was recognized above that
the $\alpha^2/r^2$ interaction was at the origin of the failure of
the perturbative analysis. Let us
define  [6] the non unitary transformation
$$\psi(r,\theta)= r^{|\alpha|} \tilde
\psi(r,\theta)$$
This is equivalent to require  that
$\psi(r,\theta)$ vanishes at the origin and that $\tilde\psi(r,\theta)$
satisfies an eigenvalue equation without the divergent
${\alpha^2\over r^2}$ singular interaction.
Indeed, the new Hamiltonian acting on $\tilde \psi(r,\theta)$ reads
\be \label{15}
2 m_o{\tilde {H}}^{\alpha}=  -\partial  _r^2  - {1\over  r}\partial  _r
-{1\over
r^2}\partial _{\theta}^2 + 2i{\alpha  \over r^2}\partial _{\theta}
-2\vert \alpha \vert {\partial_r\over r} \ee
Happily enough, the non hermitian $\vert \alpha \vert\partial_r/r$ term (in
place of the
dangerous $\alpha^2/r^2$ singular term) is now adapted for a
perturbative analysis, using the standard unperturbed Hilbert space
$\tilde \psi_{nm}^0=\psi_{nm}^0$. One finds at order $|\alpha|$ that the
correction to
the energy for the $m=0$ states is the space
integral of a total divergence, and thus only depends on the value of
the unperturbed wavefunction at the origin
\be\label{storm} {\pi\over m_0}|\psi_{n0}^0(0,\theta)|^2|\alpha|=
|\alpha|\omega \ee
Taking into account the order $\alpha$ and $|\alpha|$ corrections
(\ref{13},\ref{storm}),
one recovers
the exact spectrum (5). One also checks by explicit computation that all
higher order terms in the perturbative expansion of
(\ref{15}) are finite and  exactly cancel, as they should.
Eq. (9) means that at first order in perturbation theory,
the non hermitian vertex is equivalent to
a  ${\pi\over m_o}|\alpha|\delta(\vec r)$ interaction [6].
However, this is only true at first order in perturbation theory.
A similar computation with the non hermitian vertex
replaced by a $\delta$ interaction would lead to diverging results,   already
at
second order.
Thus claims [9] concerning this
equivalence to all orders in perturbation theory are
uncorrect\footnote{One could as well have redefined
$$\psi(r,\theta)= r^{-|\alpha|} \tilde
\psi(r,\theta)$$
 corresponding perturbatively to the self-adjoint extension where only
$J_{-\alpha}$ is retained
(one simply replaces $|\alpha|\to -|\alpha|$ in (8,9)).}.

One generalizes to $N$-anyon

\be \label{soto}  H_{N}^{\alpha}= {1\over 2m_o}\sum _{i=1}^{N}{(\vec{p}_i -
\alpha \vec A_i)^2} \ee
where $\vec A_i= \sum_{j\ne i}
{\displaystyle \vec k \times \vec r_{ij}\over \displaystyle r_{ij}^2}$
is the statistical gauge field.
If one wishes [6] to treat the  $\alpha$ and
$\alpha^2$ anyon interactions in (\ref{soto})
 as  perturbations to the free Hamiltonian
for N bosonic or fermionic particles,
the N-anyon  wave function $\psi(\vec r_1,\cdots,\vec r_N)$
(which has to vanish when $r_{ij}\to 0$)
should be rewritten as
\be\label{16} \psi(\vec r_1,\cdots,\vec r_N)= \prod_{i<j} r_{ij}^{{|\alpha|
}}
 \tilde{\psi}(\vec r_1,\cdots,\vec r_N)\ee
All the 2-body singular terms
disappear in the new Hamiltonian acting
on $\tilde{\psi}$
\be\label{17}  \tilde {H_N}^{\alpha}=\sum _{i=1}^{N}({\vec{p}_i ^2\over
2m_o}+{i\alpha\over m_o}\sum_{j\ne i}{\vec k \times \vec r_{ij}\over
r^2_{ij}}\vec \partial_i
-{\vert \alpha\vert\over m_o}\sum_{j\ne i}{\vec r_{ij}
\over r^2_{ij}}\vec \partial_i). \ee
As  a bonus, 3-body interactions, which  are not
singular, have also disappeared.
This non hermitian Hamiltonian has been used to compute at second order the
equation of state
of an anyon gas [10], and at all orders the equation of state of an anyon gas
in a strong magnetic field [11], in a second quantized formalism.

At (and only at)
first order in $|\alpha|$, the non hermitian $|\alpha|$ term
can be replaced [6] by a sum of
$\delta(\vec {r}_{ij}) $ interactions
\be\label{siti}  \tilde {H_N}^{\alpha}=\sum _{i=1}^{N}({\vec{p}_i ^2\over
2m_o}+{i\alpha\over m_o}\sum_{j\ne i}{\vec k \times \vec r_{ij}\over r^2_{ij}}
\vec \partial_i
+{\vert \alpha\vert}\sum_{j\ne i}{\pi\over m_o}\delta(\vec r_{ij}))\ee
where one has simply taken the hermitian part of
$$-{\vec r_{ij}
\over r^2_{ij}}\vec \partial_i \to {1\over 2}(-{\vec r_{ij}
\over r^2_{ij}}\vec \partial_i +
\vec \partial_i {\vec r_{ij}
\over r^2_{ij}})=
\pi\delta(\vec r_{ij})$$

ii) the perturbative approach around a good Hilbert space [7] :

Let us come back to the original problem (\ref{1}).
The standard perturbative analysis  around the standard $\alpha=0$ Hilbert
space
$\langle\psi_{n0}^0|H^\alpha-H^0|\psi_{p0}^0\rangle$ is uncorrect
due to the diverging
$\langle\psi_{n0}^0|{\alpha^2\over r^2}|\psi_{p0}^0\rangle$.
However, it is legal to developp around the unperturbed Hilbert space
$\psi_{nm}^{\alpha_o}$,
 where $\alpha_o$ should not be an integer,
otherwise $\psi_{nm=\alpha_o}^{\alpha_o}$
vanishes at the origin. Let us concentrate on the
$m=0$ states : at first order,
$E_{n0}^{\alpha_{o}}=(2n+1+|\alpha_o|)\omega$ is corrected by
\be\label{8} \langle\psi_{n0}^{\alpha_o}|H^{\alpha_o+\alpha}-H^{\alpha_o}|
            \psi_{n0}^{\alpha_o}\rangle=
            \alpha{\alpha_o\over|\alpha_o|}\omega \ee
One checks that the order $\alpha^2$ vanishes. It is certain that
the higher order corrections also vanish since (\ref{8}) co\"\i ncides
with the exact spectrum
$(2n+1+|\alpha_o+\alpha|)\omega$
provided that $\alpha\alpha_o>0$ or $|\alpha|<|\alpha_o|$ (the case
$\alpha=-\alpha_o$ is critical since the particle is only
harmonically attracted at the origin).

What has just been done is quite formal, however the limit
$\alpha_o\to 0, \alpha\alpha_o>0$ should yield a perturbative expansion
for $\alpha$ close to zero.
Indeed, in this limit, the perturbative spectrum
yields the exact spectrum
\be (2n+1)\omega+|\alpha|\omega \ee
We will come back  to this point later.

iii) the perturbative approach around a standard Hilbert space but with the
addition
of a repulsive $\delta$ interaction [8] :

Instead of the Hamiltonian (\ref{1}) one considers [12]
\be\label{un}  H'^{\alpha}=H^{\alpha}+{\pi|\alpha|\over m_o}\delta(\vec r)\ee
and instead of (\ref{soto}) one considers [13]
\be\label{deux} {H'_{N}}^{\alpha}={H_{N}}^{\alpha}
+\sum_{i<j}{2\pi|\alpha|\over m_o}\delta(\vec r_{ij})\ee
The ad-hoc introduction of repulsive $\delta$
interactions\footnote{We stress
that this $\delta$ interaction added to $H$ has nothing to do with the first
order $\delta$ introduced in (\ref{siti}) in place of the non hermitian term in
$\tilde{H}$.}
has been advocated
in the A-B context for the first order perturbative computation of the
diffusion amplitude
[12],
and in the anyon context for the second order perturbative computation
of the equation of state [13]. In both cases, the language is field theoretical
(second quantized), meaning that the first quantized $\delta$ interaction
materializes
in
a quartic $\phi^4$ type interaction. The parameter
$\pi|\alpha|/m_o$ which multiplies
the quartic interaction is choosen by hand such that perturbative divergences
cancel.
In the A-B case, the perturbative result
agrees with the first order expansion of the exact diffusion amplitude [1],
and in the
 anyon case, with the non hermitian second order perturbative equation of
state [10].
However,
the $\delta$ interaction and the $\alpha^2\over r^2$ interaction
being both ultraviolet divergent at second order, one
has to introduce a cut-off in momentum space in order to regularize them.
Then one shows that these divergences cancel in the limit of the cutoff
going to
infinity\footnote{This is different from the non hermitian approach
which, as we saw above, is correctly
defined near $\alpha=0$ and consequently does not necessitate any short
distance regulator.}.

Now the question is : why these different approaches are equivalent?

Let us first come back [7] to the $\alpha_o\to 0$ limit in the approach ii) and
focus on the matrix elements
\be \langle\psi_{n0}^{\alpha_o}|H^{\alpha_o+\alpha}|
    \psi_{p0}^{\alpha_o}\rangle =
    \langle\psi_{n0}^{\alpha_o}|H^{\alpha}|\psi_{p0}^{\alpha_o}\rangle
    +\langle\psi_{n0}^{\alpha_o}|{2\alpha\alpha_o+\alpha_o^2\over
    2m_or^2}|\psi_{p0}^{\alpha_o}\rangle \ee
In the limit $\alpha_o\to 0$
the $1/r^2$ matrix element yields a correction
$|\alpha|\omega$ which coincides with the matrix element
of a repulsive contact interaction
\be |\alpha|\omega=\langle\psi_{n0}^0|{\pi|\alpha|\over m_o}
    \delta(\vec r)|\psi_{p0}^0 \rangle\ee
This is not an accident : from the behavior near the origin
$\psi_{n0}^{\alpha_o *}\psi_{p0}^{\alpha_o}
\propto r^{2|\alpha_o|}$ one gets
\be \lim_{\alpha_o\to0,\alpha\alpha_o>0}
    {2\alpha\alpha_o+\alpha_o^2\over2m_or^2} r^{2|\alpha_o|}
    ={\pi|\alpha|\over m_o}\delta(\vec r) \ee
and thus the formal limit
\be\label{9} \langle\psi_{n0}^{\alpha_o}|H^{\alpha_o+\alpha}|
            \psi_{p0}^{\alpha_o}\rangle \to_{\alpha_o\to 0}
            \langle\psi_{n0}^0|H^\alpha+{\pi|\alpha|\over m_o}\delta(\vec r)|
            \psi_{p0}^0\rangle \ee
Qualitatively, the contact  interaction makes the flux tube impenetrable.
Of course it can be ignored if  the eigenstates vanish at the origine,
outside the subspace $m=0$ or near $\alpha=\alpha_o$ non integer.
{}From this analysis we
conclude that
the Hamiltonian
$$H'^{\alpha}=H^{\alpha}+{\pi|\alpha|\over m_o}\delta(\vec r)$$
is indeed valid for a perturbative expansion around a standard Hilbert space.
This is a formal justification of the $\delta$ interaction introduced
in an ad hoc way in the Hamiltonian (16) of the approach iii).

It remains to be shown why the approaches i) and iii) lead to identical
perturbative results.
In fact the non  unitary transformation contains the
contact interaction we are looking for [7].
If one agrees that the correct perturbative Hamiltonian is indeed
$\tilde{H}^{\alpha}$,
with the standard measure
$\int d^2\vec r {\bar {\tilde {\phi}}} {\tilde {\psi}}$,
as checked by explicit computation of the equation of state
of an anyon gas (again there are no short distance perturbative divergences,
the second virial coefficient
is exactely reproduced and the perturbative third virial coefficient [10]
is in
agreement with numerical
Monte Carlo analysis [14], the equation of state of an
anyon gas in a strong magnetic field can be computed to
 all order in perturbation theory [11]),
 what is
the correct hermitian Hamiltonian corresponding to it?
One has simply to start from
$\tilde{H}^{\alpha}$ and to perform the inverse
transformation
$$\tilde \psi(r,\theta)= r^{-|\alpha|}
\psi(r,\theta)$$
This non unitary transformation should be interpreted as a change in the
 measure
\be \int {d^2\vec r\over r^{2|\alpha|}} \ \psi^*\phi=\int d^2 \vec r \
    {\tilde\psi}^*\tilde\phi \ee
Also, and contrary to the derivation of $\tilde {H}^{\alpha}$ from
 $H^{\alpha}$ in (\ref{15}),
care has to be taken of contact singularities, meaning that the action
of the Laplacian on  $r^{|\alpha|}=e^{|\alpha|\ln r}$ has to be understood
as a distribution.
This produces the contact term
$\Delta |\alpha|\ln r=|\alpha|2\pi\delta(\vec r)$
which gives
\be r^{-|\alpha|}\left(H^{\alpha}+{|\alpha|\pi\over m_o}\delta(\vec r)\right)
    r^{|\alpha|} =\tilde{ H}^{\alpha} \ee
Thus one has deduced the Hamiltonian (16) of Ref. [8] from the non hermitian
Hamiltonian (8) of Ref. [6]. Generalization to the $N$-anyon case is
straightforward, i.e (17) is deduced from (12) through the inverse of the
non unitary transformation (11).

In conclusion, it is not surprising that the
Hamiltonians $H'^{\alpha}=H^{\alpha}+{|\alpha|\pi\over m_o}\delta(\vec r)$
and $\tilde{H}^{\alpha}$ {\it yield two perturbative approaches which are
identical.}
However, as already stressed above,
perturbative short distance
singularities are absent in the non
hermitian Hamiltonian approach, (as well  as
3-body
interactions in the $N$-anyon case (12)), thus the economy of a short distance
regulator.

\vfill\eject
{\bf {References}\rm}

1. Y. Aharonov and D. Bohm, Phys. Rev. {\bf 115} (1959) 485

2 J. M. Leinaas and J. Myrheim,  Nuovo  Cimento B {\bf 37}  (1977)   1


3. R. Jackiw, in "M.A.B. Beg Memorial Volume", World scientific 1991 ; C.
Manuel and R. Tarrach,
Phys.
Lett. B {\bf 268} (1991) 222; J. Grundberg, T.H. Hansson, A. Karlhede and J.M.
Leinaas, Mod. Phys. Lett. B {\bf 5} (1991) 539

4. Y. Aharonov, C. K. Au, E. C. Lerner and J. Q. Liang, Phys. Rev. D {\bf 29}
   (1984)    2396

5. A. Comtet, Y. Georgelin and S. Ouvry, J. Phys. A : Math Gen. {\bf
22} (1989) 3917

6.  J. McCabe and S. Ouvry, Phys. Lett. B {\bf 260} 113 (1991) 13;
   A. Comtet, J. McCabe and S. Ouvry, Phys. Lett. B {\bf 260} (1991) 372;
   D. Sen, Nucl. Phys. B {\bf 360} (1991) 397

7. A. Dasni\a`eres de Veigy and S. Ouvry, "De la Diffusion
 	Aharonov- Bohm" Compte Rendus de l'Acad\'emie des
	Sciences , {\bf 318} (1994) 19

8.  O. Bergman and G. Lozano, Massachusetts preprint (1993); D. Z. Freedman,
G. Lozano and N. Rius; CTP preprint 2216 (1993);
	M. A. Valle Basagoiti, Phys. Lett. B{\bf 306} (1993) 307;
    R. Emparan and M. A. Valle Basagoiti, Mod. Phys. Lett. A{\bf 8} (1993)
3291;
  see also G. Amelino-Camelia, MIT-CTP-2242 preprint; see also
C. Manuel and R. Tarrach, UB-ECM-PF 19/93 preprint

9. see D. Sen in [5]; D. Sen and R. Chitra, Phys. Rev. B {\bf 45} (1991) 881;
in the presence of a strong B field see
J.Y.Kim et al, Phys. Lett. B {\bf 306} (1993) 91 and
Phys. Rev. D {\bf 48} (1993) 4839; for a detailed discussion of the non
equivalence see
G. Amelino-Camelia, Phys. Lett. B {\bf 286} (1992) 97; for an exact derivation
of
the equation of state of an anyon gas in a strong B-field to all order
in perturbation theory see [11]

10.  A. Dasni\a`eres de Veigy and S. Ouvry, Phys. Lett. B {\bf 291} (1992)130;
A. Dasni\a`eres de Veigy and S. Ouvry, Nucl. Phys. B[FS] {\bf 388}, 715
      (1992)

11.  A. Dasni\`eres de Veigy and S. Ouvry,
Phys. Rev. Lett. {\bf 72} (1994) 600

12. see O. Bergman et al in [8]

13. see R. Emparan et al in [8]

14.  J. Myrheim and K. Olaussen, Phys. Lett. B {\bf 299} (1993) 267

\end{document}